\documentclass[prd,11pt]{revtex4-1}

\usepackage{physics}
\usepackage{bm}
\usepackage{amsmath}    
\usepackage{graphicx}   
\usepackage{verbatim}   
\usepackage{color}      
\usepackage{subfigure}  
\usepackage{hyperref}   
\definecolor{amaranth}{rgb}{0.9, 0.17, 0.31}
\definecolor{purple(munsell)}{rgb}{0.62, 0.0, 0.77}
\definecolor{americanrose}{rgb}{1.0, 0.01, 0.24}
\definecolor{palatinateblue}{rgb}{0.15, 0.23, 0.89}
\definecolor{royalblue(web)}{rgb}{0.25, 0.41, 0.88}
\definecolor{hanpurple}{rgb}{0.32, 0.09, 0.98}
\definecolor{beaublue}{rgb}{0.74, 0.83, 0.9}
\definecolor{carminered}{rgb}{1.0, 0.0, 0.22}
\definecolor{brightpink}{rgb}{1.0, 0.0, 0.5}
\definecolor{vividviolet}{rgb}{0.62, 0.0, 1.0}

\hypersetup{ linktoc=all,
    colorlinks, linkcolor={palatinateblue},
    citecolor={brightpink}, urlcolor={amaranth}}

\newcommand{\be}{\begin{equation}}
\newcommand{\ee}{\end{equation}}
\newcommand{\bs}{\begin{split}} 
\newcommand{\bea}{\begin{eqnarray}}
\newcommand{\eea}{\end{eqnarray}}

\begin{document}

\title{\Large Rigidity and Parallelism in the spacetime }

\author{Nosratollah Jafari}\email{nosrat.jafari@gmail.com }

\affiliation{Department of Physics, Nazarbayev University,\\Kabanbay Batyr Ave 53, Nur-Sultan, 010000, Kazakhstan.}

\begin{abstract}

The effect of the linear-fractional transformations on the parallel lines in the spacetime has been studied. Fock-
Lorentz transformations maps a line to a line, from which one can obtain the combinations rule for
the velocities in the Fock-Lorentz transformations. Rigidity is defined as a consequences of holding
parallelism under the transformations. The Fock-Lorentz transformations do not preserve rigidity,
which leads to some novel results such as growing distances alongside with advancing time. Also,
it is shown that the time coordinates of events will come closer to each other in the transformed
coordinates by going back in time

\vspace{13cm}

\noindent Keywords: Linear-Fractional Transformations, Fock-Lorentz Transformations,
Rigidity, Parallelism.

\end{abstract}

\maketitle

\tableofcontents


\newpage

\section{Introduction}

\par The Fock-Lorentz transformations, which are linear-fractional generalization
of the Lorentz transformations, have been obtained by V. A. Fock  \cite{Foc}.
Other authors have obtained similar transformations before or after the V. A. Fock.
A complete of list of references could be found in \cite{man1}.
Properties of the Fock-Lorentz transformations and their relations with other group
theoretical extensions of the special theory of relativity with two fundamental
constants have been studied in \cite{Guo1, Guo2, Huan1, Huan2 }.

\par Studying the geometrical properties of the linear-fractional transformations,
and especially the Fock-Lorentz transformations, can help us to have a better
visualization of these transformations.  These transformations act on the spacetime
coordinates, thus we will consider a line or two parallel lines in the spacetime
for studying the geometrical properties of the linear-fractional transformations.
However, we can assume a straight line or two parallel lines in the true physical
space $(x,y,z)$ which in fact can lead to an expanding universe. There are some
studies which have related the expansion of the universe with this property of
the linear-fractional transformations and the Fock-Lorentz transformations
\cite{man1, step1, step2}.

\par Mapping a line under the Lorentz or the the Fock-Lorentz transformations or any
desired transformations in the sacetime can specify the behaviour of a free particle
under these transformations. Also, mapping two parallel lines in the spacetime
under any transformations can show a better understanding of the concept of
simultaneous evens and the behavior of these evevts under these transformations.
We will assign a number to the non-rigidity. The sign of this number can show us that
these lines will come closer or go away from each other under the transformations.
Also, these lines will remain parallel to each other if this number takes
a zero value, which is happened obviously for the Lorentz transformations.

\section{ Parallel motions in the spacetime }
\subsection{ The Lorentz Transformation case}
Taking $c = 1$, the standard Lorentz transformation in the spacetime are
\be \label{eq1}
   \left\{\begin{array}{cl}  x'=\gamma ( x - ut)\cr
   t'=\gamma ( t - u x), \end{array}\right.
\ee
Consider a free particle $ x = vt + x_0 $ in the $S$ frame or $(xt)$ plane.
The Lorentz transformation maps this line to
\be x' = v't' + \frac{x_0}{\gamma(1-uv)}, \ee
where
\be v' = \frac{v - u}{1 -vu}, \ee
is the velocity of the particle in the $S'$ frame.  Note that, this is the Einstein's
combination rule for velocities.  Now we consider two particles in a parallel motion
in $(xt)$ plane, by which we mean the velocity of the two particles are the same.
\be  \label{eq3}
  \left\{\begin{array}{cl}  x =vt + x_0 \cr
   \tilde{x}= v t + \tilde{x}_0. \end{array}\right.
\ee
The Lorentz transformation maps these parallel lines to
\be  \label{eq4}
   \left\{\begin{array}{cl} x' = v't' + \frac{x_0}{\gamma(1-uv)} \cr
   \tilde{x}' = v't' + \frac{ \tilde{x}_0 }{\gamma(1-uv)} , \end{array}\right.
\ee
which are also parallel lines in the $(x't')$ plane.
Note that the Galilean transformations will map Eq.(\ref{eq3}) to
\be  \label{eq5}
  \left\{\begin{array}{cl}  x' = v't' + x_0 \cr
   \tilde{ x}' = v't' + \tilde {x}_0 , \end{array}\right.
\ee
where $ v'= v-u $. The main difference, apart from the relativistic addition of
velocities, is the factor $ \gamma^{-1}(1 - uv)^{-1}$ multiplying the initial
positions.

\subsection{Fock- Lorentz Transformations case}
The Fock-Lorentz transformation in (1+1) dimensions is
\be  \label{eq 7}
  \left\{\begin{array}{cl}
   x'=
   \displaystyle\frac{\gamma( x - ut)}{ 1+ \lambda u \gamma x - \lambda (\gamma -1)t }
   \cr
   t'=
   \displaystyle\frac{\gamma(t-u x)}{1+\lambda u \gamma x - \lambda (\gamma -1)t },
  \end{array}\right.
\ee
where  $c=1$, and $\lambda$ is a  very small constant with the dimension of
$\textrm{TL}^{-2}$.
These transformations will map the free particle $x = vt + x_0 $ to
\be \label{eq 8}
 x' = \left[\frac{ v - u - \lambda x_0(\gamma -1 )/\gamma}{ 1- uv + \lambda u x_0}
 \right] t' + \frac{x_0}{\gamma( 1- uv + \lambda u x_0 ) },
\ee
which is also the equation of a line in the $(x't')$ plane. Thus, the Fock-Lorentz
transformations map a free particle motion to a free particle motion.
As in the Lorentz case, we can interpret the slope of this line
\be \label{eq 9}
  v' =  \frac{ v - u - \lambda x_0(\gamma -1 )/\gamma}{ 1- uv + \lambda u x_0},
\ee
as the  composition rule for the velocities in the Fock-Lorentz transformations.
This velocity depends on the initial position of the particle $x_0$. Note also that
to obtain (\ref{eq 8}) we have assumed a free particle with constat velocity in the
$(xt)$. We will find a more general expression for the combination rule of the
velocities in the next section.

\par The Fock-Lorentz transformations will map two free particles which are in
parallel motions in the $S$ frame
\be  \label{eq 10}
 \left\{\begin{array}{cl}  x= vt+ x_0 \cr
 \tilde{x}= vt+ \tilde{x}_0, \end{array}\right.
\ee
to
\be  \label{eq 11}
 \left\{\begin{array}{cl}
 x' = \left[\frac{v -u- \lambda x_0(\gamma -1)/\gamma}{1- uv +\lambda u x_0}\right] t'
     + \frac{x_0}{\gamma( 1- uv +\lambda u x_0 ) } \cr
 \tilde{x}' = \left[
  \frac{v - u - \lambda \tilde{x}_0 (\gamma -1)/\gamma}{1- uv + \lambda u  \tilde{x}_0}  \right] t' +
 \frac{ \tilde{x}_0}{\gamma( 1- uv + \lambda u \tilde{x}_0 ) }, \end{array}\right.
\ee
Although , the velocities of these two particle are the same in $S$ frame,
their velocities are not the same in the $S'$ frame.  In other words,
the Fock-Lorentz transformations do not preserve parallelism.
 The lines in the $(x't')$ plane, Eq.~(\ref{eq 11}) will cross each other at
\be  \label{eq 12}
  \left(t'_\lambda, x'_\lambda\right) =
  \left(\frac{-1}{\lambda}, \frac{\lambda + u - v}{\lambda[1- uv + \lambda u x_0]}
  \right).
\ee
The time component $t'_\lambda$ does not depend on $x_0$ or $\tilde{x}_0$,
which shows that the crossing points are on a line, parallel to the $x'$ axis.
 We have obtained  Eq.~(\ref{eq 12})
for $\gamma \gg 1$, but in general the value of $t'_\lambda$ is independent of the
$x_0$ or $\tilde{x}_0$.  The general expressions for $(t'_\lambda, x'_\lambda)$
are more complex and we only give the value of $t'_\lambda$, which is
\be
  t'_\lambda= \frac{-1}{\lambda} \cdot \frac{1- uv}{  (\gamma-1 )/ \gamma - uv}.
\ee

\section{linear- fractional transformations in general form}
We consider a general form for the linear-fractional transformations in
$(1+1)$ dimensions from the $S$  frame with coordinates $(x,t)$ to the $S'$ frame
with coordinates $(x',t')$.
\be  \label{eq 14}
 x' = \frac{a_1 + A_{11}x + A_{10}t}{1 + \alpha_1x + \alpha_0t}
, \qquad t' =\frac{a_0 + A_{01}x + A_{00}t}{1 + \alpha_1x + \alpha_0t}.
\ee
Here,  $A_{\mu\nu}$ is a general $2\times 2$ invertible matrix and $\alpha_0$ and
$ \alpha_1$ are constants.

\par Taking derivatives of  Eq.~(\ref{eq 14}) we obtain the velocity
$v'= dx'/dt'$ of the particle in the $S'$ frame  as
\be   \label{eq 15}
 v'= \frac{(A_{11} - \alpha_1a_1)v + (A_{10} - \alpha_0a_1)
 + (A_{10}\alpha_1 - A_{11}\alpha_0)(x - vt)}
 {(A_{01} - \alpha_1a_0)v + (A_{00} - \alpha_0 a_0)
 + (A_{00}\alpha_1 - A_{01}\alpha_0)(x - vt)}.
\ee
This formulas can be rewritten in another compact form
\be  \label{eq 16}
v'= \frac{A_{11}v + A_{10} - x'(\alpha_1 v + \alpha_0)}
{ A_{01}v + A_{00} - t' (\alpha_1 v + \alpha_0)}.
\ee
We should mention that this formula has been obtained intially by Kerner
\cite{kern}. Two expressions for the velocity in  Eq.~(\ref{eq 15}) and
Eq.~(\ref{eq 16}) are equivalent; they differ only in dependency
of the velocity of the particle to the $S$ frame coordinates or the $S'$  frame
coordinates. It is enough to put the expressions for $x'$ and $t'$ from
Eq.~(\ref{eq 14}) in the Eq.~(\ref{eq 16}) to reach to the Eq.~(\ref{eq 15}).

\subsection{Velocity in Fock-Lorentz case}
For the Fock-Lorentz transformations we have
\be  \label{eq 19}
[A_{\mu\nu}]= \left(\begin{array}{cc} \gamma & -\gamma u \\
 -\gamma u & \gamma \\  \end{array} \right),
\ee
where
\be \label{eq 20}
 \alpha_1= \lambda u \gamma, \qquad \alpha_0= - \lambda (\gamma -1).
\ee
Substituting these values for $A_{\mu\nu}$, $\alpha_1$, and $\alpha_0$  in
Eq.~(\ref{eq 16}) we obtain the velocity of the particle in the $S'$ frame as
\be
  v' =\frac{v - u + \lambda \left(\gamma^{-1}-1\right)(x -vt)}
  {1 - uv + \lambda u(x - vt)} .
\ee
This is a general formula for the velocity.  If we take $x = x_0+vt $ for a free
particle with constant speed, we get the same result as in Eq.~(\ref{eq 9}) of
section 2.2.  Also, from Eq.~(\ref{eq 16}) we get
\be
 v'= \frac{  v -u - \lambda x' \left( uvx + \gamma^{-1} -1\right)}
 {1- uv -  \lambda t' \left( uvx +\gamma^{-1} -1\right) } ,
\ee
which explicitly shows the dependency of $v'$ on the $(x',t')$ coordinates of the
particle in the $S'$ frame.

\par We have obtained the Einstein's combination rule for velocities from the slope
of the line in the transformed spacetime.  We repeated the same procedure for the
Fock-Lorentz transformations and obtain the combination rule for the velocities.
However, the difference between these two combination rules is that in the
Fock-Lorentz case the combination rule depends on the initial position of the
particle.  This is a well known matter, but we have obtained the formulas from mapping
a line in the spacetime and this method can lead us to a better visualization.
Similar expressions have been obtained initially by Kerner \cite{kern},
by Manida \cite{man1}, and Stepanov \cite{step1}.

\section{Rigidity in the linear-fractional transformations of the spacetime}
Using the general forms of transformations in Eq.~(\ref{eq 14}) we can compute
the difference
\be
 x'- \tilde{x}' = \frac{a_1 + A_{11}x + A_{10}t}{1 + \alpha_1x + \alpha_0t}
                - \frac{a_1 + A_{11} \tilde{x} + A_{10} \tilde{t} }{1 + \alpha_1
                \tilde{x} + \alpha_0 \tilde{t} },
\ee
which can be simplified to the first order of $\alpha$ to
\bea
 x' - \tilde{x}' \simeq
  \Big[(A_{11} - a_1\alpha_1)(x- \tilde{x} ) + (A_{10} - a_1\alpha_0)(t- \tilde{t}) + \nonumber \cr
(A_{11}\alpha_0 - A_{10}\alpha_1)(x \tilde{t}- \tilde{x} t)\Big]
\Big[1 - \alpha_1(x + \tilde{x}) - \alpha_0(t +\tilde{t})\Big] + O(\alpha2).
\eea
For time component the difference will be
\be \label{eq 32} t'- \tilde{t}'  =\frac{a_0 + A_{01}x + A_{00}t}{1 + \alpha_1x +
\alpha_0t} -  \frac{a_0 + A_{01}\tilde{x} + A_{00}\tilde{t} }
{1 + \alpha_1 \tilde{x} + \alpha_0\tilde{t} }, \ee
or in the simplified form
\bea
t' - \tilde{t}'= \Big[(A_{01} - a_0\alpha_1)(x - \tilde{x}) + (A_{00} - a_0\alpha_0)(t-
\tilde{t}) +  \nonumber \\
(A_{00}\alpha_1 - A_{01}\alpha_0)(x \tilde{t}- \tilde{x}t)\Big]   \Big[ 1 - \alpha_1(x
+ \tilde{x}) - \alpha_0(t + \tilde{t}) \Big] + O(\alpha^2). \eea
Here, for any convenient choices of the $[A_{\mu\nu}]$ and also $\alpha_0$,
$\alpha_1$ , $a_0$ $a_1$ we can define some new transformations in the spacetime
for which we can check that these transformations will preserve the rigidity or not.

\subsection{Fock-Lorentz transformations case and its visualization }
For the specific case of the Fock-Lorentz transformations
\be  \label{eq 34}
x' - \tilde{x}' = \frac{ \gamma \Big[ 1  + \lambda t (\gamma -1 )/ \gamma \Big]  }
 { \Big[ 1 + \lambda u
\gamma x  - \lambda ( \gamma- 1)t \Big]  \Big[ 1 + \lambda u \gamma \tilde{x}  -
\lambda ( \gamma- 1) \tilde{t} \Big]  } (x - \tilde{x})  ,
\ee
which in the first order of $ \lambda $ can be simplified as
\be  \label{eq 35} x' - \tilde{x}' =  \gamma \Big[ 1 - \lambda u
\gamma (x + \tilde{x})
+  \lambda \frac{( \gamma- 1)(2 \gamma +1 )}{\gamma } t \Big] (x - \tilde{x}),
\ee
 also in the second equation we have put $t= \tilde{t}$.

\par We have obtained the contraction of the length in the Fock-Lorentz
transformations, which has smaller amount with respect to the Lorentz case.
If we take $ x - \tilde{x} = L $ we will see that $L'$ in the transomed
coordinates will depends on the coordinates of the rod in $(xt)$ plane.
In other words, the coordinates of the beginning $\tilde{x}$ and the end
$x$ of the a rod in  $(xt)$ plane and also the time of observation will
be entered in this formals.
In the Lorentz case, we have assumed the same parallelogram in the $(xt)$ plane and
we study the deformations of this parallelogram under the the Fock-Lorentz
transformations.
We define a new function $f(X,\tilde{X} )$ by taking
\be  \label{eq 36}
 f(X,\tilde{X} ) =
 1- \sigma (x + \tilde{x}) + \rho t,
\ee
in which
\be   \label{eq 37}
\sigma= \lambda \gamma u, \qquad
\rho=  \lambda \frac{( \gamma- 1)(2 \gamma +1 )}{\gamma }.
\ee
Thus,  Eq.~(\ref{eq 35}) can be rewritten as
\be \label{eq 38}  x' - \tilde{x}'=\gamma  f(X,\tilde{X} )(x - \tilde{x}) .\ee
As indicted two particles are moving with the same velocity in the $(xt)$ plane
and by putting the values of $ x$ and $ \tilde{x}$
in the expression for $ f(X,\tilde{X} )$  we can relate
$ f(X,\tilde{X} )$ to the $ f(X_0,\tilde{X}_0 )$ by
\be  \label{eq 39} f(X,\tilde{X} )=  f(X_0,\tilde{X}_0 ) + g(t, t_0), \ee
in which the function $ g(t, t_0) $ indicates their difference
\be  g (t, t_0) = (\rho - 2 \sigma v )(t- t_0).\ee
For high velocities $ \gamma \gg 1 $  we have  approximately
\be \label{eq 41} g (t, t_0 ) = 2 \lambda \gamma ( 1 - uv ) ( t- t_0), \ee  which is
obviously greater than zero. Thus,
\be  f(X,\tilde{X} )\gg  f(X_0,\tilde{X}_0 ) ,\ee
which shows that the  $ x'_0 - \tilde{x}'_0$  interval will grow
under the motion of these two particles. Thus, the rigidity is not preserved under the Fock-Lorentz
transformations. In fact, going back in time, the $ x'_0 - \tilde{x}'_0$
interval will be shrunk and will go to the zero in $ t = - 1/\lambda c^2 $,
which is easily seen from Eq.(\ref{eq 35}) by putting the numerator of the
right hand side of this equation to zero.

\par If we want to be precise,  $ x'_0 - \tilde{x}'_0 $ can be rewritten in terms of
$(x_0 - \tilde{x}_0)$ in the compact form similar to the Eq.~(\ref{eq 38}) as
\be \label{eq }
 x'_0 - \tilde{x}'_0 =\gamma  f(X_0,\tilde{X}_0 )(x_0 - \tilde{x}_0) .
\ee
In the $(x,t)$ plane we have $x - \tilde{x}= x_0 - \tilde{x}_0 $, thus by using of the
Eq(\ref{eq 36}) and Eq.(\ref{eq 39})
we can assign a value $N_{nr}$ for the non-rigidity
\be
 N_{nr}:= (x' - \tilde{x}') -(x_0 - \tilde{x}_0)
        =  \gamma  g(t,t_0)(x_0 - \tilde{x}_0 ).
 \ee
For $\gamma \gg 1$ by using of the value of $g(t, t_0) $  from
Eq.~(\ref{eq 41}) we find
\be \label{eq 45}
 N_{nr} = 2\lambda \gamma ( 1 - uv ) ( t- t_0)(x_0 - \tilde{x}_0 ),
\ee
which is a positive value as mentioned in the above and will be increased
by going forward in time or for the bigger value of $t-t_0$.
The  negativity or positivity  of the value of the rigidity $ N_{nr}$ will
specify that the parallel line in the $(x't')$ will come close together or not.

On the other hand, for $ t'- \tilde{t}'$ from Eq.(\ref{eq 32})
we have
\be   \label {eq 43}
 t' - \tilde{t}' = \frac {- u \gamma (  1+ \lambda t ) }
  { \Big[ 1 + \lambda u \gamma x  - \lambda ( \gamma- 1)t \Big]
  \Big[ 1 + \lambda u \gamma \tilde{x}  - \lambda ( \gamma- 1) \tilde{t} \Big]  }
  (x - \tilde{x}) ,
\ee
which  can be simplified in first order of $ \lambda$ as
\be  \label{eq 44}
 t' - \tilde{t}' = - \gamma u \Big[ 1 - \lambda u \gamma ( x +\tilde{x})
 + \lambda ( 2\gamma- 1)t \Big] (x - \tilde{x}),
\ee
also in the second equation we have put $t= \tilde{t}$.
Here, as in the above we introduce
\be  \Theta( X, \tilde{X} )= 1 - \sigma (x + \tilde{x} ) + \beta t ,  \ee
in which $\sigma$ is given by Eq.~(\ref{eq 37}) and $ \beta= \lambda ( 2 \gamma -1)$.
Thus, we can rewrite Eq.~(\ref{eq 44}) as
\be t' - \tilde{t}' =  - u \gamma \Theta( X, \tilde{X} )  (x - \tilde{x}) .\ee
By putting the equations of motions of the assumed two parallel particles
we can relate
$\Theta( X, \tilde{X} ) $ to the $\Theta( X_0, \tilde{X}_0 ) $ as
\be \Theta( X, \tilde{X} )= \Theta( X_0, \tilde{X}_0 ) + \chi (t, t_0), \ee
in which
\be  \chi (t, t_0)= (\beta - 2 \sigma v )( t- t_0) .\ee
These  equations can help us to compare the $  t' - \tilde{t}'$ with
$  t'_0 - \tilde{t}'_0$.
For high velocities $ \gamma \gg 1 $, we find
\be \chi (t, t_0)= 2 \lambda \gamma (1-uv )( t- t_0) ,\ee
which has a positive value. Thus,
\be \left\vert t' - \tilde{t}'\right\vert >
\left\vert t'_0 - \tilde{t}'_0\right\vert,
\ee
which shows that the value of $|t'_0 - \tilde{t}'_0 |$
will grow by motions of these two particles. In other words,
by going back in the time the $|t' - \tilde{t}' |$ will become smaller and in the
$t= -1/\lambda c^2 $ will go to zero as seen from
Eq.~(\ref{eq 43}) by putting the numerator of the right side of this equation to zero.

It is also possible to define the rigidity as preserving
$ t- \tilde{t}=  t_0- \tilde{t}_0 $ under the
spacetime transformations which will be brought to
$ t'- \tilde{t}'=  t'_0- \tilde{t}'_0 $ in the $(x't')$
spacetime. Also,  we should take the same initial positions for the particles which
have started from different initial times. The Lorentz transformations are
symmetric for the time and positions coordinates, and this definition does not
lead to significant different aspects. But, the Fock-Lorentz transformations are
not symmetric for the time and positions coordinates, and we can encounter with
different aspects of the non-rigidity. But, the $t'_\lambda$ line and crossing of
the the particles paths on $t'_\lambda$ line will be occurred as in the same way.
We postpone this study to the future studies.

\section{Conclusions}
The Lorentz transformations are affine transformations which map a straight line
in the spacetime to a straight line. On the other hand, the affine transformations
will preserve the parallelism of the lines and the spacetime is rigid under these
transformations. Thus, we can say that the value of non-rigidity is zero for the
affine transformations.

\par In \cite{jaf}, it was shown that the Fock-Lorentz transformations in
the spactime and also the MS transformation in the energy- momentum space
as projective transformations.  The projective transformations also map a
straight line to a straight line, but they do not keep the parallelism.
The amount of the deviations from the parallelism can be expressed by introducing
the rigidity. As mentioned, the positivity and the negativity of the non-rigidity
can show that these lines will go away from each other or come
close to each other.

\par We can do the same study for the proposed varying speed of light theories
by investigation the effects of these theories  on a line or two parallel lines
in the spacetime. Also by calculating the non-rigidity we can specify the behaviour
of these  theories in any region of the spacetime  \cite{mag1, mag2}.
It may happen that these theories attain different value for the non-rigidity and
have different aspects under mapping of the parallel lines. We can do all of these
in a very small region of spacetime for comparing the local behaviour of these
theories by the Lorentz transformations. Also, one can do the same study for the
transformations in the spacetime which has been obtained as a position space images
of the doubly special theories by some authors \cite{kim, der, hos}.

\par On the other hand, Maguejo-Smolin transformations are linear-fractional
transformations in the energy-momentum space \cite{MS1}, and the results of this
study is applicable to MS DSR, but with some cares . It should be noted that the
concept of the line in the spacetime is not the same in the energy-momentum space .

\end{document}